\PassOptionsToPackage{unicode,hypertexnames=false}{hyperref}
\PassOptionsToPackage{hyphens}{url}
\PassOptionsToPackage{dvipsnames,svgnames,x11names}{xcolor}
\documentclass[
  12pt]{article}
\usepackage{algorithm}
\usepackage{algpseudocode}
\usepackage{amsmath, amsthm, amssymb}

\newtheorem{proposition}{Proposition}
\usepackage{iftex}
\usepackage{booktabs}
\usepackage{multirow}
\usepackage{makecell}
\usepackage{booktabs}
\usepackage{enumitem}

\ifPDFTeX
  \usepackage[T1]{fontenc}
  \usepackage[utf8]{inputenc}
  \usepackage{textcomp} % provide euro and other symbols
\else % if luatex or xetex
  \usepackage{unicode-math}
  \defaultfontfeatures{Scale=MatchLowercase}
  \defaultfontfeatures[\rmfamily]{Ligatures=TeX,Scale=1}
\fi
\usepackage{lmodern}
\ifPDFTeX\else  
\fi
\IfFileExists{upquote.sty}{\usepackage{upquote}}{}
\IfFileExists{microtype.sty}{% use microtype if available
  \usepackage[]{microtype}
  \UseMicrotypeSet[protrusion]{basicmath} % disable protrusion for tt fonts
}{}
\makeatletter
\@ifundefined{KOMAClassName}{% if non-KOMA class
  \IfFileExists{parskip.sty}{%
    \usepackage{parskip}
  }{% else
    \setlength{\parindent}{0pt}
    \setlength{\parskip}{6pt plus 2pt minus 1pt}}
}{% if KOMA class
  \KOMAoptions{parskip=half}}
\makeatother
\usepackage[dvipsnames]{xcolor}
\usepackage[leftbars,color]{changebar}
\cbcolor{ForestGreen}
\makeatletter
\ifx\paragraph\undefined\else
  \let\oldparagraph\paragraph
  \renewcommand{\paragraph}{
    \@ifstar
      \xxxParagraphStar
      \xxxParagraphNoStar
  }
  \newcommand{\xxxParagraphStar}[1]{\oldparagraph*{#1}\mbox{}}
  \newcommand{\xxxParagraphNoStar}[1]{\oldparagraph{#1}\mbox{}}
\fi
\ifx\subparagraph\undefined\else
  \let\oldsubparagraph\subparagraph
  \renewcommand{\subparagraph}{
    \@ifstar
      \xxxSubParagraphStar
      \xxxSubParagraphNoStar
  }
  \newcommand{\xxxSubParagraphStar}[1]{\oldsubparagraph*{#1}\mbox{}}
  \newcommand{\xxxSubParagraphNoStar}[1]{\oldsubparagraph{#1}\mbox{}}
\fi
\makeatother

\usepackage{longtable,booktabs,array}
\usepackage{calc} % for calculating minipage widths
\usepackage{etoolbox}
\makeatletter
\patchcmd\longtable{\par}{\if@noskipsec\mbox{}\fi\par}{}{}
\makeatother
\IfFileExists{footnotehyper.sty}{\usepackage{footnotehyper}}{\usepackage{footnote}}
\makesavenoteenv{longtable}
\usepackage{graphicx}
\makeatletter
\def\maxwidth{\ifdim\Gin@nat@width>\linewidth\linewidth\else\Gin@nat@width\fi}
\def\maxheight{\ifdim\Gin@nat@height>\textheight\textheight\else\Gin@nat@height\fi}
\makeatother
\setkeys{Gin}{width=\maxwidth,height=\maxheight,keepaspectratio}
\makeatletter
\def\fps@figure{htbp}
\makeatother

\makeatletter
\@ifpackageloaded{caption}{}{\usepackage{caption}}
\AtBeginDocument{%
\ifdefined\contentsname
  \renewcommand*\contentsname{Table of contents}
\else
  \newcommand\contentsname{Table of contents}
\fi
\ifdefined\listfigurename
  \renewcommand*\listfigurename{List of Figures}
\else
  \newcommand\listfigurename{List of Figures}
\fi
\ifdefined\listtablename
  \renewcommand*\listtablename{List of Tables}
\else
  \newcommand\listtablename{List of Tables}
\fi
\ifdefined\figurename
  \renewcommand*\figurename{Figure}
\else
  \newcommand\figurename{Figure}
\fi
\ifdefined\tablename
  \renewcommand*\tablename{Table}
\else
  \newcommand\tablename{Table}
\fi
}
\@ifpackageloaded{float}{}{\usepackage{float}}
\floatstyle{ruled}
\@ifundefined{c@chapter}{\newfloat{codelisting}{h}{lop}}{\newfloat{codelisting}{h}{lop}[chapter]}
\floatname{codelisting}{Listing}

\makeatother
\makeatletter
\@ifpackageloaded{caption}{}{\usepackage{caption}}
\@ifpackageloaded{subcaption}{}{\usepackage{subcaption}}
\makeatother

\usepackage[leftbars,color]{changebar}
\cbcolor{red}
\ifLuaTeX
  \usepackage{selnolig}  % disable illegal ligatures
\fi
\usepackage[]{natbib}
\usepackage{xr-hyper}
\usepackage{bookmark}
\usepackage{tabularx}
\usepackage{booktabs}

\IfFileExists{xurl.sty}{\usepackage{xurl}}{} % add URL line breaks if available
\hypersetup{
  pdftitle={Multiple change-point detection via bottom-up scanning},
  pdfkeywords={Frequent change-points, Nonparametrics, Graph-based methods,
               Hierarchical merging, High-dimensional data},
  colorlinks=true,
  linkcolor={blue},
  filecolor={Maroon},
  citecolor={Blue},
  urlcolor={Blue},
  pdfcreator={LaTeX via pandoc}}

\newcommand{\anon}{1}

\if1\anon
  \hypersetup{pdfauthor={Jinhyeok Park; Taejoon Yang; Hyeyoung Maeng; Hoseung Song}}
\fi
\begin{document}

\def\spacingset#1{\renewcommand{\baselinestretch}%
{#1}\small\normalsize} \spacingset{1}

%%%%%%%%%%%%%%%%%%%%%%%%%%%%%%%%%%%%%%%%%%%%%%%%%%%%%%%%%%%%%%%%%%%%%%%%%%%%%%

\if1\anon
{
	\title{\bf Multiple change-point detection via bottom-up scanning}
	\author{Jinhyeok Park$^{*}$\\
		Department of Industrial and Systems Engineering, KAIST.\\
		Taejoon Yang$^{*}$\\
		Department of Industrial and Systems Engineering, KAIST.\\
		Hyeyoung Maeng$^{\dagger}$\\
		Department of Statistics, Ewha Womans University.\\
		and \\
		Hoseung Song$^{\dagger\ddagger}$\\
		Department of Industrial and Systems Engineering, KAIST.}
	\maketitle
	
	\begingroup
	\renewcommand{\thefootnote}{\fnsymbol{footnote}}
	\footnotetext[1]{Equal contribution.}
	\footnotetext[2]{Corresponding authors.}
	\footnotetext[3]{This work was supported, in part, by the National Research Foundation of Korea (NRF) grant (RS-2026-25471551, RS-2026-25517529) and the National IT Industry Promotion Agency (NIPA) grant (RS-2026-25621650), both funded by the Korea government (MSIT).}
	\endgroup
} \fi

\if0\anon
{
  \bigskip
  \bigskip
  \bigskip
  \begin{center}
    {\LARGE\bf Multiple change-point detection via bottom-up scanning}
\end{center}
  \medskip
} \fi

\bigskip
\begin{abstract}
We study nonparametric multiple change-point detection for high-dimensional sequences, aiming to identify time points at which the underlying distribution changes. While many existing methods perform well when change-points are well-separated, their performance can deteriorate when structural breaks are densely clustered. To address this challenge, we propose gBottomup, a graph-based bottom-up framework for multiple change-point detection in high-dimensional settings. gBottomup constructs a hierarchical segmentation by proposing merges of adjacent segments and verifying them through an unmerge rule that combines absolute significance with relative local heterogeneity, thereby adaptively refining partitions and estimating the number of change-points. Simulation results demonstrate that gBottomup performs reliably across a range of structural configurations and is particularly effective in frequent change-point settings, where existing top-down procedures may lose sensitivity. We illustrate the proposed method through an analysis of high-dimensional sensor data from an experimental hydraulic system. 
\end{abstract}

\noindent%
\vfill

\newpage
\spacingset{1.8} % DON'T change the spacing!

%%%%%%%%%%%%%%%%%%%%%%%%%%%%%%%%%%%%%%%%%%%%%%%%%%%%%%%%%%%%%%%%%%%%%%%%%%%%%%

\section{Introduction}\label{sec:intro}\label{sec:topdown}

Change-point detection aims to identify distributional changes in data and estimate their locations. Such problems arise in a wide range of application domains, including medical condition monitoring \citep{bosc2003automatic,staudacher2005new,malladi2013online}, genomics \citep{zhang2010dna, jiang2015codex}, finance \citep{kim2022kwicss}, and social networks \citep{hazrati2021nsocial}. For example, engineering monitoring systems generate high-dimensional sensor records whose distributions can change as operating conditions evolve. When changes occur in rapid succession, identifying each transition requires sensitivity to short regimes as well as flexibility regarding the form of the change. The hydraulic-system application in Section~\ref{sec:real_data} illustrates this challenge through repeated load cycles with changing component conditions.

We consider offline multiple change-point detection for a sequence $\{X_t\}_{t=1}^T\subset\mathbb{R}^d$. The observations satisfy
\begin{equation}\label{eq:model}
X_t\sim F_j,\qquad \tau_j<t\le\tau_{j+1},\qquad j=0,\ldots,m,
\end{equation}
where $0=\tau_0<\tau_1<\cdots<\tau_m<\tau_{m+1}=T$ and $F_{j-1}\ne F_j$ for $j=1,\ldots,m$. Our goal is to estimate both the number of change-points, $m$, and their locations, $\mathcal T=\{\tau_1,\ldots,\tau_m\}$, without specifying a parametric family for the segment distributions.

Nonparametric approaches include kernel methods \citep{harchaoui2008kernel,arlot2019kernel,song2024practical}, empirical divergence methods such as E-Divisive \citep{matteson2014nonparametric}, and graph-based methods \citep{chen2015graph,song2022asymptotic,chen2023graph}. Graph-based procedures represent similarities among observations through a graph and assess whether its edge pattern differs across a candidate boundary. This representation accommodates complex multivariate structure and can detect changes beyond the mean.

For multiple change-point detection, the boundary statistic must be combined with a search procedure. Wild Binary Segmentation (WBS) \citep{fryzlewicz2014wild} and Seeded Binary Segmentation (SBS) \citep{kovacs2023seeded} search collections of subintervals to identify changes that may be difficult to isolate over the entire sequence. These schemes follow a top-down strategy that begins with the full sequence, identifies a candidate change-point over a large interval, and recursively applies the same procedure to the resulting sub-intervals. \citet{zhang2021multi} incorporated graph-based statistics into these frameworks, yielding gMulti(G.WBS) and gMulti(G.SBS). Nevertheless, their sensitivity can deteriorate when several changes occur within a short interval. We call these \emph{frequent change-point} settings and focus particularly on clustered changes.

Bottom-up procedures offer a complementary search strategy by starting from fine-scale segments and progressively merging neighboring regions. The Tail-Greedy Unbalanced Haar transform targets piecewise-constant signals \citep{fryzlewicz2018tail}, TrendSegment extends this approach to piecewise-linear signals \citep{maeng2024detecting}, and BUHDA addresses high-dimensional mean changes \citep{maeng2026high}. These methods focus on specified signal features, such as means or trends. Our aim is to retain the local sensitivity of bottom-up search while targeting broader distributional changes.

We propose gBottomup, a graph-based bottom-up framework built on the generalized edge-count statistic of \citet{chu2019asymptotic}. The procedure has three phases. Local Merging screens adjacent segments from a fine initial partition; Hierarchical Merging then generates candidate boundaries through score-based merges; and Model Selection prunes the candidates using an early-stopped BIC-type criterion. Both merging phases verify provisional merges through an unmerge rule that preserves a boundary when either its absolute evidence of a change is sufficient or its heterogeneity exceeds that of its local surroundings. This combination is designed to protect nearby changes during candidate generation while allowing subsequent removal of spurious boundaries.

\begin{figure}[h]
\centering
\includegraphics[width=0.92\linewidth]{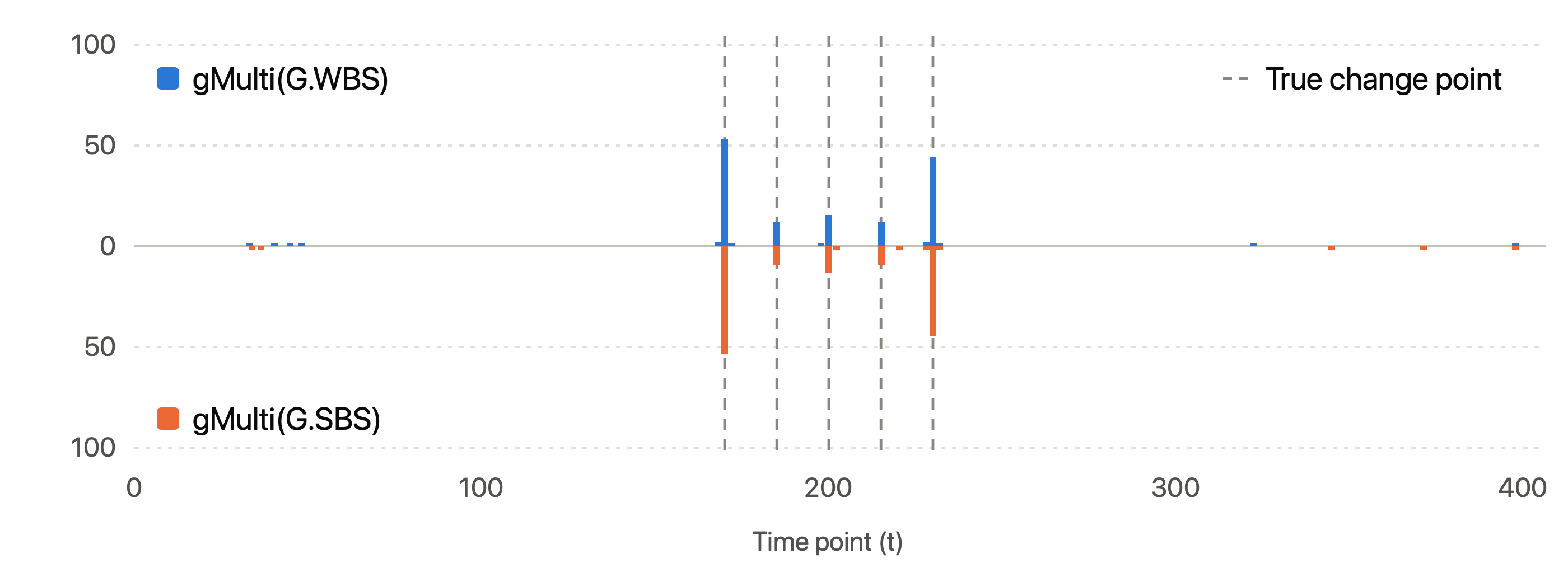}
\par\smallskip
\includegraphics[width=0.92\linewidth]{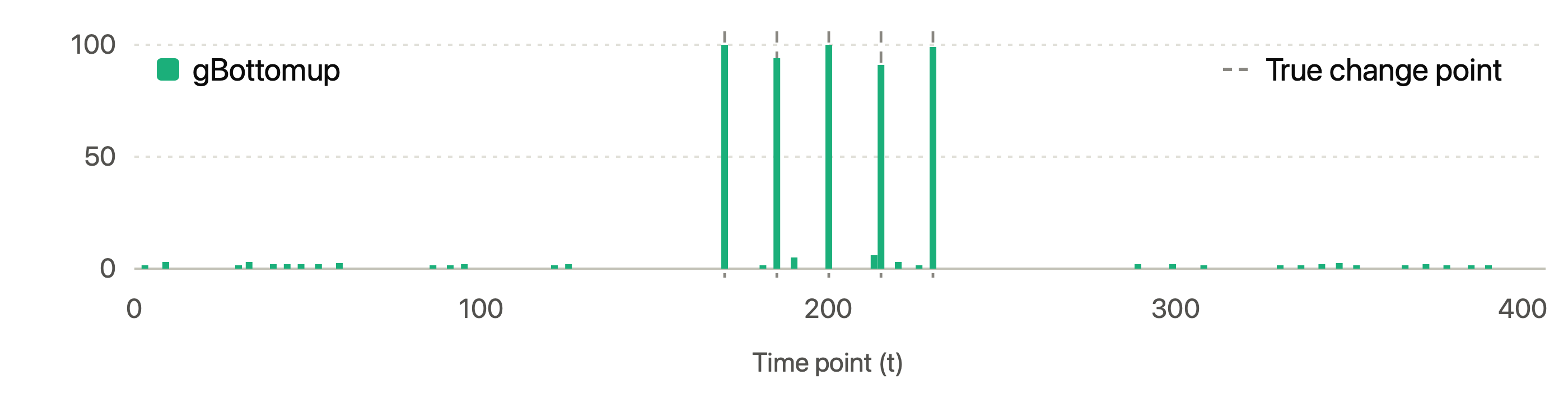}
\caption{Detection frequencies for gMulti(G.WBS), gMulti(G.SBS), and gBottomup under the frequent mean-change setting of Model~5 ($T=400$, $d=500$, 100 replications). Dashed lines mark the true change-points at $\mathcal T=\{170,185,200,215,230\}$.}
\label{fig:frequent_comparison}\label{fig:sbs_dense}\label{fig:gbus_dense}
\end{figure}

Figure~\ref{fig:frequent_comparison} illustrates the motivation using the frequent mean-change configuration of Section~\ref{sec:na_setup}. The graph-based top-down methods detect some changes near the edges of the cluster but have substantially lower sensitivity to its interior changes. gBottomup maintains high detection frequencies across all five boundaries. The broader simulations assess this advantage across spacing configurations and types of distributional change, while runtime experiments compare computational performance.

The remainder of this paper is organized as follows. Section~\ref{sec:method} describes the algorithm and its properties, and Section~\ref{sec:choice_of_param} gives the tuning choices. Section~\ref{sec:analysis} presents the simulation study, followed by the hydraulic-system application in Section~\ref{sec:real_data} and a discussion in Section~\ref{sec-conc}. Code is available at \url{https://github.com/zinhyeok/gBottomup}.

%%%%%%%%%%%%%%%%%%%%%%%%%%%%%%%%%%%%%%%%%%%%%%%%%%%%%%%%%%%%%%%%%%%%%%%%%%%%%%

\section{Proposed method}\label{sec:method}

\subsection{Overview and graph-based scores}\label{sec:overview}

gBottomup starts from a fine partition and proceeds through three phases. Local Merging screens disjoint adjacent pairs, Hierarchical Merging builds a candidate hierarchy by merging low-scoring pairs, and Model Selection prunes the resulting candidates. Both merging phases use the same unmerge rule to retain boundaries supported by either absolute or relative evidence of a change. Algorithm~\ref{alg:driver} summarizes the full procedure of gBottomup.
\begin{algorithm}[h]
	\small
	\caption{gBottomup}\label{alg:driver}
	\begin{algorithmic}[1]
		\Require $\{X_t\}_{t=1}^T$; segment length $L$; ratios $\rho_{local},\rho_{merge}$; penalty constant $c$; significance level $\alpha$
		\Ensure estimated change-point set $\hat\tau$
		\State Partition the sequence into $N=\lfloor T/L\rfloor$\footnotemark segments of length $L$.
		\State \textbf{Phase 1:} Form disjoint adjacent pairs and apply Unmerge (Algorithm~\ref{alg:unmerge}) for $h_{local}$ levels.
		\State \textbf{Phase 2:} Rank adjacent boundaries, merge greedily, and apply Unmerge; complete the hierarchy and extract $\tilde\tau$ by~\eqref{eq:candidates}.
		\State \textbf{Phase 3:} Apply early-stopped backward elimination to $\tilde\tau$ using~\eqref{eq:epmbic}.
		\State \Return $\hat\tau$
	\end{algorithmic}
\end{algorithm}
\footnotetext{$\lfloor x \rfloor$ denotes the largest integer that is no larger than $x$.}

%%%%%%%%%%%%%%%%%%%%%%%%% footnote 글씨 안나옴 %%%%%%%%%%%%%%%%%%%%%%%%%

For observations on an interval $[a,b]$, let $G^{[a,b]}$ be a similarity graph. We use a $k$-MST, the union of $k$ edge-disjoint minimum spanning trees \citep{chen2015graph,chu2019asymptotic}. At a candidate split $t$, let $\mathbf R^{[a,b]}(t)=(R_1^{[a,b]}(t),R_2^{[a,b]}(t))^\top$ count edges with both endpoints in $[a,t]$ and $[t+1,b]$, respectively. The generalized edge-count statistic is
\begin{equation}\label{eq:graphscore}
S^{[a,b]}(t)=
\left(\mathbf R^{[a,b]}(t)-\mathrm E\{\mathbf R^{[a,b]}(t)\}\right)^\top
\left(\Sigma^{[a,b]}(t)\right)^{-1}
\left(\mathbf R^{[a,b]}(t)-\mathrm E\{\mathbf R^{[a,b]}(t)\}\right),
\end{equation}
where $\Sigma^{[a,b]}(t)=\operatorname{Var}(\mathbf R^{[a,b]}(t))$ and the moments are evaluated under the permutation null. Large values provide evidence of a distributional change. For a fixed split, the statistic has an asymptotic $\chi^2_2$ null distribution under suitable regularity conditions \citep{chu2019asymptotic,zhu2024limiting}. The graph parameter $k$ is specified in Section~\ref{sec:choice_of_param}.

For adjacent segments $B=[a,\nu]$ and $B'=[\nu+1,b]$, write $W=\{B,B'\}$ and $S(W)=S^{[a,b]}(\nu)$. We also write $W_\nu$ to identify its internal boundary. Unlike a scan over every split in an interval, the algorithm evaluates $S$ only at boundaries of the current partition.

%%%%%%%%%%%%%%%%%%%%%%%%%%%%%%

\subsection{Phase 1: Local Merging}\label{sec:phase1}

\begin{figure}[h] 
	\centering
	\includegraphics[width=\linewidth]{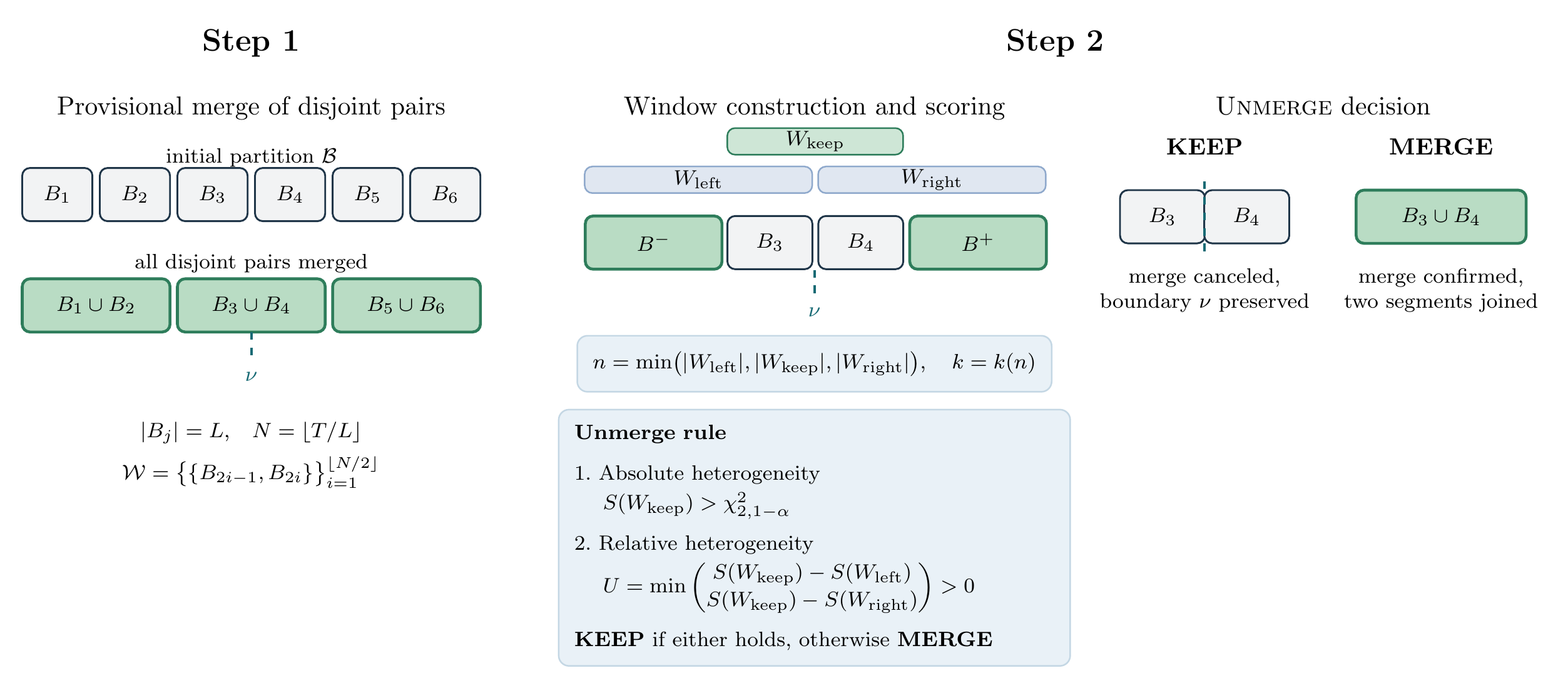}
	\caption{Illustration of Phase 1, Local Merging.} 
	\label{fig:phase1}
\end{figure}

Let $\mathcal B=\{B_1,\ldots,B_N\}$ be the initial partition, with $N=\lfloor T/L\rfloor$ and segment length $L=5$, following \citet{zhang2021multi}. Smaller $L$ improves resolution but can produce unstable local scores. At each level, if the current partition has $s$ segments, we provisionally merge the disjoint pairs
$\mathcal W=\{\{B_{2i-1},B_{2i}\}:i=1,\ldots,\lfloor s/2\rfloor\}$.
The unmerge rule then verifies all pairs using this provisionally merged partition; a schematic is given in Figure \ref{fig:phase1}.

For a pair $\{B_L,B_R\}\in\mathcal W$, let $B^-$ and $B^+$ denote the segments immediately to the left and right of the merged block. Define
\[
W_{\mathrm{left}}=\{B^-,B_L\},\qquad
W_{\mathrm{keep}}=\{B_L,B_R\},\qquad
W_{\mathrm{right}}=\{B_R,B^+\}.
\]
A neighboring window is omitted when the corresponding segment does not exist. To keep the graph construction consistent across the compared windows, we use a shared $k$ determined by
\begin{equation}\label{eq:n_phase1}
n=\min\{|W_{\mathrm{left}}|,|W_{\mathrm{keep}}|,|W_{\mathrm{right}}|\},
\end{equation}
where $|W|$ is the number of observations in $W$ and undefined windows are omitted. The rule mapping $n$ to $k$ is given in~\eqref{para:k}.

A provisional merge is canceled (\texttt{KEEP}) if either of two criteria holds. The \emph{absolute heterogeneity} criterion is $S(W_{\mathrm{keep}})>\chi^2_{2,1-\alpha}$. The \emph{relative heterogeneity} criterion is $U>0$, where
\begin{equation}
\label{eq:criterion2}
U=\begin{cases}
\min\bigl(S(W_{\mathrm{keep}})-S(W_{\mathrm{left}}),\ S(W_{\mathrm{keep}})-S(W_{\mathrm{right}})\bigr)
& \text{if both neighbors exist,}\\
S(W_{\mathrm{keep}})-S(W_{\mathrm{left}})
& \text{if only $W_{\mathrm{left}}$ exists,}\\
S(W_{\mathrm{keep}})-S(W_{\mathrm{right}})
& \text{if only $W_{\mathrm{right}}$ exists.}
\end{cases}
\end{equation}
Otherwise the pair remains merged (\texttt{MERGE}). All decisions are made before updating the partition, as specified in Algorithm~\ref{alg:unmerge}.

The relative criterion serves a similar purpose to the adjust pass of \citet{maeng2026high}, which restores a boundary that is more heterogeneous than its surroundings. The absolute criterion retains a boundary with an above-threshold keep-window score even when neighboring windows also show strong evidence of change. We retain a boundary whenever either criterion holds because a boundary removed at this stage cannot be recovered later, whereas retained spurious candidates can be pruned in Phase 3. Our rule also differs from the adjust pass by canceling only the provisional merge, without merging the restored segments with their outer neighbors.

\begin{algorithm}[h]
\scriptsize
	\caption{Unmerge}
	\label{alg:unmerge}
	\begin{algorithmic}[1]
		\Procedure{Unmerge}{$\mathcal{B}, \mathcal{W}$}
		\State $\mathcal{R} \gets \emptyset$
		\For{$\{B_L, B_R\} \in \mathcal{W}$} \Comment{$\mathcal{B}$ is not updated within this loop}
		    \State $B^{-} \gets$ segment left of $B_L$ in $\mathcal{B}$, if it exists 
            \State $B^{+} \gets$ segment right of $B_R$ in $\mathcal{B}$, if it exists
		    \State $W_{\mathrm{keep}} \gets \{B_L, B_R\}$; \;
		           $W_{\mathrm{left}} \gets \{B^{-}, B_L\}$ if $B^{-}$ exists; \;
		           $W_{\mathrm{right}} \gets \{B_R, B^{+}\}$ if $B^{+}$ exists
		    \State $n \gets \min(|W| : W$ exists$)$, \; $k \gets k(n)$
		           \Comment{Eq.~\eqref{eq:n_phase1}: undefined windows omitted}
		    \State compute $S(W)$ for each existing $W$ among $W_{\mathrm{keep}}, W_{\mathrm{left}}, W_{\mathrm{right}}$, using $k$
		    \State $U \gets$ Eq.~\eqref{eq:criterion2}
		    \If{$S(W_{\mathrm{keep}}) > \chi^2_{2, 1-\alpha}$ \textbf{or} $U > 0$}
		        \State \texttt{KEEP}: $\mathcal{R} \gets \mathcal{R} \cup \{\{B_L,B_R\}\}$ \Comment{boundary retained}
		    \Else
		        \State \texttt{MERGE}: the pair remains merged in $\mathcal{B}$ \Comment{segments joined}
		    \EndIf
		\EndFor
		\State $\mathcal{B} \gets \mathcal{B}$ with every pair in $\mathcal{R}$ split back
		       \Comment{$\mathcal{B}$ updated once, after the loop}
		\State \textbf{return} $\mathcal{B}, \mathcal{R}$
		\EndProcedure
	\end{algorithmic}
\end{algorithm}

The number of Local Merging levels is controlled by
\begin{align}
h_{local}&=\lfloor\rho_{local}H_{total}\rfloor,\label{equ:h_pre}\\
H_{total}&\approx\frac{\ln N}{-\ln(1-\rho_{merge})}.\label{equ:H_total}
\end{align}
Here $\rho_{local}\in(0,1)$ determines the extent of initial screening, and $\rho_{merge}\in(0,1)$ controls the fraction of low-scoring boundaries proposed for merging in Phase 2. The choice of $\rho_{local}$ and $\rho_{merge}$ is discussed in Section \ref{sec:calibration}, and the depth approximation is derived in Supplement~D.

%%%%%%%%%%%%%%%%%%%%%%%%%%%%%%

\subsection{Phase 2: Hierarchical Merging}\label{sec:phase2}

\begin{figure}[h]
\centering
\includegraphics[width=\linewidth]{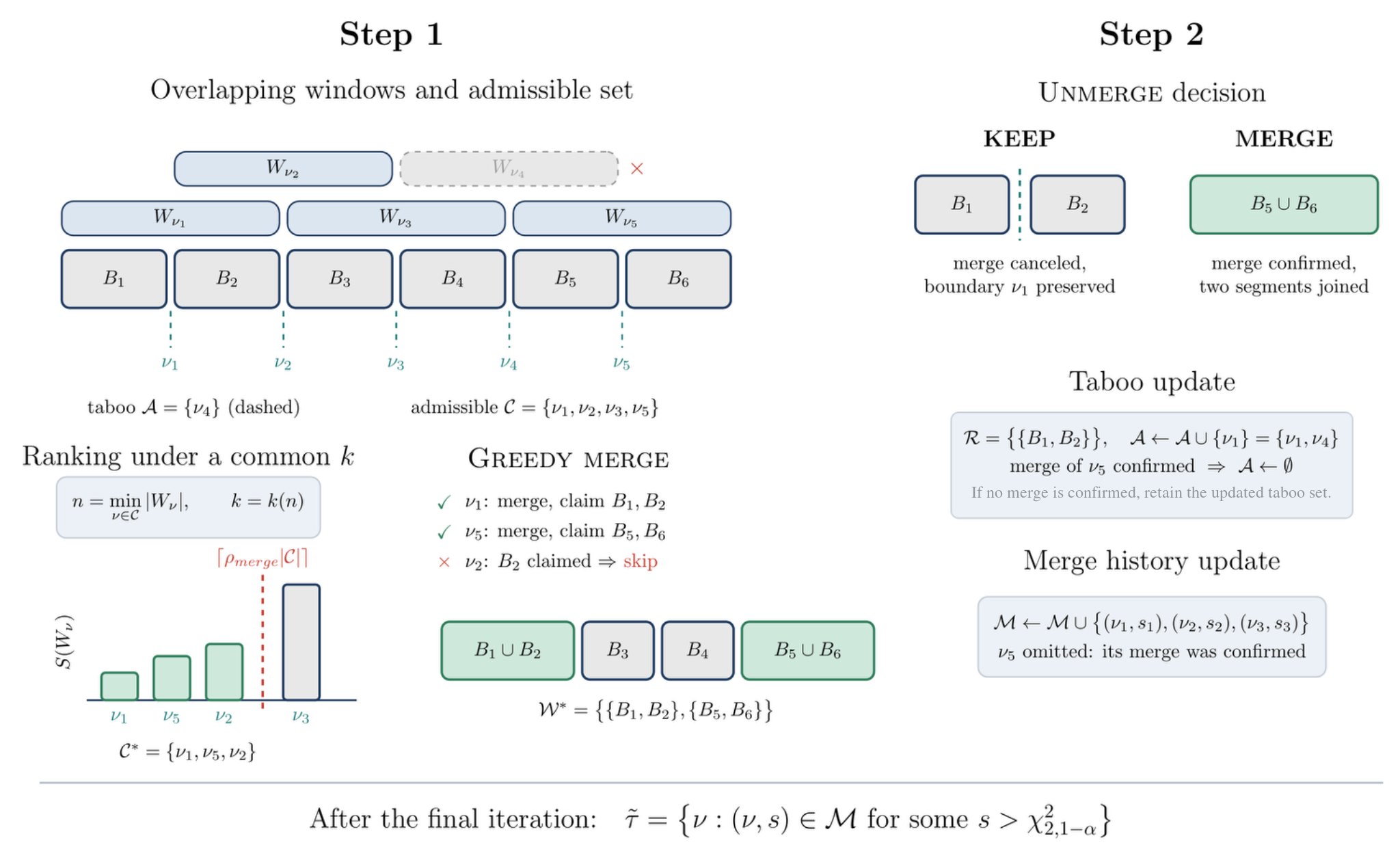}
\caption{One iteration of Hierarchical Merging: global ranking and disjoint provisional merges, followed by unmerge verification and updates to the taboo set and merge history.}
\label{fig:phase2_iteration}
\end{figure}

Phase 2 uses the same merge-and-verify structure, but proposes pairs by score rather than position. For the current partition $\mathcal B=\{B_1,\ldots,B_s\}$, each adjacent boundary $\nu_i$ has window $W_{\nu_i}=\{B_i,B_{i+1}\}$. These windows overlap, and their segment identities remain fixed within an iteration. Figure~\ref{fig:phase2_iteration} illustrates the procedure.

\textbf{Ranking and provisional merging.}
A taboo set $\mathcal A$, initially empty, excludes previously retained boundaries until the segmentation changes. The admissible set is $\mathcal C=\{\nu_i:i=1,\ldots,s-1\}\setminus\mathcal A$. All admissible windows use a shared graph parameter based on
\begin{equation}\label{eq:n_phase2}
n=\min_{\nu\in\mathcal C}|W_\nu|.
\end{equation}
These \emph{global ranking scores} use a $k$ selected across $\mathcal C$, whereas each unmerge comparison uses its own shared $k$ from~\eqref{eq:n_phase1}. We select
\begin{equation}\label{eq:merge_quantile}
\mathcal C^*=\{\nu\in\mathcal C:S(W_\nu)\le Q_{\rho_{merge}}(\{S(W_\eta):\eta\in\mathcal C\})\},
\end{equation}
where $Q_{\rho_{merge}}$ is the $\lceil\rho_{merge}|\mathcal C|\rceil$\footnote{$\lceil x \rceil$ denotes the smallest integer that is no smaller than $x$.}-th smallest score. Processing $\mathcal C^*$ in ascending score order, with ties broken by boundary index, we merge a pair only when neither segment has been claimed by an earlier pair. This produces disjoint provisional pairs $\mathcal W^*$. Details are given in Algorithm~\ref{alg:greedymerge}.
\begin{algorithm}[h]
	\scriptsize
	\caption{Greedy merge}
	\label{alg:greedymerge}
	\begin{algorithmic}[1]
		\Procedure{GreedyMerge}{$\mathcal{B}, \mathcal{C}, \rho_{merge}$}
		\State $n \gets \min_{\nu\in\mathcal{C}}|W_\nu|$
		\State $k \gets k(n)$
		\Comment{graph parameter; Eq.~\eqref{para:k}}
		\State compute $S(W_\nu)$ for every $\nu\in\mathcal{C}$ using $k$
		\Comment{Eq.~\eqref{eq:n_phase2}}
		\State $\mathcal{C}^{*}\gets\{\nu\in\mathcal{C}:S(W_\nu)\le
		Q_{\rho_{merge}}(\{S(W_\eta):\eta\in\mathcal{C}\})\}$
		\State $\mathcal{W}^{*}\gets\emptyset$, \; $\mathrm{claimed}\gets\emptyset$
		\For{$\nu_i\in\mathcal{C}^{*}$ in ascending $S(W_{\nu_i})$} \Comment{$B_i, B_{i+1}$ fixed at the start of this iteration}
		% \State $(B_{i}, B_{i+1}) \gets$ the two segments of $\mathcal{B}$ forming
		%        $W_{\nu_i}$
		\If{$B_i \notin \mathrm{claimed}$ \textbf{and} $B_{i+1} \notin \mathrm{claimed}$}
		\State $\mathcal{B} \gets \mathcal{B}$ with the consecutive pair
		$(B_i, B_{i+1})$ replaced by $B_i \cup B_{i+1}$
		\Comment{order of $\mathcal{B}$ preserved}
		\State $\mathrm{claimed}\gets\mathrm{claimed}\cup\{B_i, B_{i+1}\}$
		\State $\mathcal{W}^{*}\gets\mathcal{W}^{*}\cup\{\{B_i, B_{i+1}\}\}$
		\Else
		\State \textbf{skip} $\nu_i$
		\Comment{$B_i$ or $B_{i+1}$ already claimed by an earlier
			(lower-score) pair; not added to the taboo set}
		\EndIf
		\EndFor
		\State \Return $\mathcal{B}$, $\mathcal{W}^{*}$, $\{(\nu,S(W_\nu)):\nu\in\mathcal{C}\}$
		\EndProcedure
	\end{algorithmic}
\end{algorithm}

\textbf{Verification and taboo updates.}
Algorithm~\ref{alg:unmerge} verifies $\mathcal W^*$ and returns the retained pairs $\mathcal R$. Boundaries in $\mathcal R$ are added to $\mathcal A$; the entire taboo set is reset when at least one ordinary merge is confirmed. In Figure \ref{fig:phase2_iteration}, retaining $\nu_1$ first updates the taboo set to $\{\nu_1,\nu_4\}$; confirmation of the merge at $\nu_5$ then resets the entire set to $\varnothing$. Skipped overlapping pairs are not added to $\mathcal A$. If no admissible boundary remains before a single segment is obtained, a terminal \emph{force merge} stage suspends unmerge verification and the taboo restriction, repeatedly removing the weakest remaining boundaries until the hierarchy is complete.

\textbf{Candidate extraction.}
During ordinary and force-merge iterations, we record a multiset $\mathcal M$ of boundary-score pairs $(\nu,s)$, where $s$ is the global ranking score. At each iteration, scores of all admissible boundaries are recorded except those whose merges are confirmed in that iteration. Thus a surviving boundary may appear repeatedly with different scores. Phase 1 scores are not recorded because their short windows may give unreliable scores. The candidate set is
\begin{equation}\label{eq:candidates}
\tilde\tau=\{\nu:(\nu,s)\in\mathcal M\text{ for some }s>\chi^2_{2,1-\alpha}\}.
\end{equation}
Retaining boundaries that exceed the threshold at any recorded scale is intended to preserve sensitivity; Phase 3 then removes unnecessary candidates. The full procedure of Phase 2 is summarized in
Algorithm~\ref{alg:gbus_merge}.
\begin{algorithm}[h]
	\scriptsize
	\caption{Hierarchical Merging (Phase 2)}
	\label{alg:gbus_merge}
	\begin{algorithmic}[1]
		\Procedure{HierarchicalMerge}{$\mathcal{B},\rho_{merge},\alpha$}
		\State $\mathcal{A}\gets\emptyset$, \; $\mathcal{M}\gets\emptyset$
		\Comment{taboo set; merge history}
		\While{$|\mathcal{B}|>1$}
		\State $\mathcal{C}\gets\{\nu_i : i=1,\ldots,|\mathcal{B}|-1\}\setminus\mathcal{A}$;
		\textbf{ if }$\mathcal{C}=\emptyset$\textbf{ then break}
		\Comment{no admissible boundary $\to$ force merge}
		\State $(\mathcal{B}, \mathcal{W}^{*}, \{(\nu,S(W_\nu)):\nu\in\mathcal{C}\})\gets\Call{GreedyMerge}{\mathcal{B},\mathcal{C},\rho_{merge}}$
		\Comment{Algorithm~\ref{alg:greedymerge}}
		\State $(\mathcal{B}, \mathcal{R})\gets\Call{Unmerge}{\mathcal{B}, \mathcal{W}^{*}}$
		\Comment{Algorithm~\ref{alg:unmerge}}
		\State $\mathcal{A}\gets\mathcal{A}\cup\{\nu : W_\nu\in\mathcal{R}\}$
		\State $\mathcal{M}.\mathrm{append}\bigl(\{(\nu,S(W_\nu)):\nu\in\mathcal{C},\;
		\nu\notin\{\eta:W_\eta\in\mathcal{W}^{*}\setminus\mathcal{R}\}\}\bigr)$
		\Comment{surviving boundaries}
		\State \textbf{ if }$\mathcal{W}^{*}\setminus\mathcal{R}\neq\emptyset$\textbf{ then }$\mathcal{A}\gets\emptyset$
		\Comment{taboo reset (ordinary merges only)}
		\EndWhile
		\While{$|\mathcal{B}|>1$}
		\Comment{\textbf{Force merge}: terminal; unmerge and $\mathcal{A}$ suspended}
		\State $\mathcal{C}\gets\{\nu_i : i=1,\ldots,|\mathcal{B}|-1\}$
		\State $(\mathcal{B}, \mathcal{W}^{*}, \{(\nu,S(W_\nu)):\nu\in\mathcal{C}\})\gets\Call{GreedyMerge}{\mathcal{B},\mathcal{C},\rho_{merge}}$
		\State $\mathcal{M}.\mathrm{append}\bigl(\{(\nu,S(W_\nu)):\nu\in\mathcal{C},\;
		\nu\notin\{\eta:W_\eta\in\mathcal{W}^{*}\}\}\bigr)$
		\EndWhile
		\State \textbf{return} $\tilde{\tau}\gets
		\{\nu : (\nu, s)\in\mathcal{M}
		\text{ for some } s > \chi^2_{2,1-\alpha}\}$
		\EndProcedure
	\end{algorithmic}
\end{algorithm}

%%%%%%%%%%%%%%%%%%%%%%%%%%%%%%

\subsection{Phase 3: Model Selection}\label{sec:modelselection}

Let $\tilde\tau=\{\tilde\tau_1,\ldots,\tilde\tau_{\tilde m}\}$, with $\tilde\tau_0=0$ and $\tilde\tau_{\tilde m+1}=T$. Following \citet{zhang2021multi}, define the expanded Adjacent Sum statistic
\begin{equation}\label{eq:eas}
eAS(\tilde\tau)=\sum_{j=1}^{\tilde m}S(W_{\tilde\tau_j}),
\qquad
W_{\tilde\tau_j}=\{[\tilde\tau_{j-1}+1,\tilde\tau_j],[\tilde\tau_j+1,\tilde\tau_{j+1}]\}.
\end{equation}
Each term measures evidence at a candidate boundary. Because spurious candidates can also contribute positive scores, we balance this sum against an extended pseudo-modified BIC, adapted from \citet{zhang2007modified}:
\begin{equation}\label{eq:epmbic}
\operatorname{ep\text{-}mBIC}(\tilde\tau)
=eAS(\tilde\tau)-\sum_{j=1}^{\tilde m+1}\log(\tilde\tau_j-\tilde\tau_{j-1})-c\tilde m\log T.
\end{equation}
The penalty constant $c$ is calibrated empirically as described in Section~\ref{sec:choice_of_param}.

Penalty-based criteria can underestimate the number of frequent changes \citep{fryzlewicz2020detecting}. We therefore use Early-Stopped Backward Elimination (ES-BE): at each step, delete the candidate whose removal yields the largest ep-mBIC, but stop immediately if that deletion does not improve the current criterion. ES-BE provides conservative pruning relative to selection over the same complete greedy deletion path, as established in Section \ref{sec:theory}. Algorithm \ref{alg:be} gives the complete procedure.
\begin{algorithm}[h]
	\small
	\caption{Early-Stopped Backward Elimination (ES-BE) with ep-mBIC}
	\label{alg:be}
	\begin{algorithmic}[1]
		\Procedure{ES-BE}{$\tilde{\tau}$}
		\State $\hat{\tau} \gets \tilde{\tau}$
		\While{$|\hat{\tau}| > 0$}
		\State $\tau^{*} \gets \operatorname*{argmax}_{\tau \in \hat{\tau}}\,
		\text{ep-mBIC}(\hat{\tau} \setminus \{\tau\})$
		\State \textbf{if} $\text{ep-mBIC}(\hat{\tau} \setminus \{\tau^{*}\}) \le
		\text{ep-mBIC}(\hat{\tau})$ \textbf{then break}
		\Comment{early stop: no improvement}
		\State $\hat{\tau} \gets \hat{\tau} \setminus \{\tau^{*}\}$
		\Comment{remove; model improves}
		\EndWhile
		\State \textbf{return} $\hat{\tau}$
		\EndProcedure
	\end{algorithmic}
\end{algorithm}

%%%%%%%%%%%%%%%%%%%%%%%%%%%%%%

\subsection{Properties of the algorithm}\label{sec:theory}

The taboo mechanism controls repeated verification, while early stopping provides a conservative pruning rule relative to complete-path selection. We formalize these two properties below.

\begin{proposition}
    Suppose that Phase 2 starts with $N_2$ segments and $\rho_{merge}\in(0,1)$. The procedure terminates after finitely many iterations. Ordinary merging performs at most $N_2(N_2-1)/2$ unmerge pair verifications. Counting repeated evaluations, the total number of graph-score evaluations in Phase 2, including force merging, is $O(N_2^2/\rho_{merge})$, regardless of which provisional merges are confirmed.
\end{proposition}

The bound holds without assuming that a fixed fraction of proposed merges is confirmed. Computational cost also depends on window size: under bounded $k$, evaluating a window of $w$ observations in $\mathbb R^d$ costs $O{w^2(d+\log w)}$. Supplement B accounts for these costs and reports empirical runtimes.

\textbf{Conservativeness of early stopping.} Let $\widehat\tau_{full}$ be any ep-mBIC maximizer along the complete greedy backward-elimination path starting from $\widetilde\tau$, with the same tie choices as ES-BE along their shared path. Then
\begin{align*}
    \widehat\tau_{full}\subseteq\widehat\tau,
\end{align*}
where $\widehat\tau$ is the ES-BE output. Thus ES-BE can retain additional candidates but cannot discard a candidate retained by selection over that complete path. Proofs of both properties are given in Supplement A.

%%%%%%%%%%%%%%%%%%%%%%%%%%%%%%%%%%%%%%%%%%%%%%%%%%%%%%%%%%%%%%%%%%%%%%%%%%%%%%%%%%%%%%%%%%%%%%%%%%%%%%%%%%%%%%%%%%%%%%%%

\section{Choice of tuning parameters}\label{sec:choice_of_param}
\label{sec:choice_k}

We use $L=5$ and $\alpha=0.05$. Following \citet{zhang2021multi}, the graph parameter is chosen as
\begin{equation}\label{para:k}
k=\begin{cases}
\min(30,\lfloor\sqrt n\rfloor),&\text{in Phases 1 and 2},\\
\min(5,\lfloor\sqrt n\rfloor),&\text{in Phase 3}.
\end{cases}
\end{equation}
For Phases 1 and 2, $n$ is the shared minimum window size in~\eqref{eq:n_phase1} or~\eqref{eq:n_phase2}. In Phase 3, evaluation at $\tilde\tau_j$ uses $n=\tilde\tau_{j+1}-\tilde\tau_{j-1}$. The cap of 30 limits graph-construction cost on large windows; the cap of 5 keeps $k$ constant in Phase 3 except on very short windows. Using the shared minimum in the merging phases keeps the graph parameter common across scores compared in the same decision.

\label{sec:calibration}
We choose $c=2$, $\rho_{local}=0.3$, and $\rho_{merge}=0.2$ through a calibration study separate from the evaluations in Sections~\ref{sec:analysis} and~\ref{sec:real_data}. The study crosses mean and scale changes with well-separated and frequent configurations, each with five change-points and $d\in\{20,50,100,500,1000\}$. We first select $c$ and then calibrate $(\rho_{local},\rho_{merge})$ jointly. The selected values balance true and false discoveries across the four calibration models and are used throughout the numerical studies. Full designs, results, and parameter trade-offs are provided in Supplement~C.

%%%%%%%%%%%%%%%%%%%%%%%%%%%%%%%%%%%%%%%%%%%%%%%%%%%%%%%%%%%%%%%%%%%%%%%%%%%%%%%%%%%%%%%%%%%%%%%%%%%%%%%%%%%%%%%%%%%%%%%%

\section{Numerical analysis}\label{sec:analysis}

We evaluate detection and segmentation performance under three distributional scenarios and three spacing configurations. Runtime comparisons are reported in Supplement~B.

%%%%%%%%%%%%%%%%%%%%%%%%%%%%%%

\subsection{Simulation setup}\label{sec:na_setup}

We set $d=500$ and use the gBottomup settings from Section~\ref{sec:choice_of_param}: $L=5$, $c=2$, $\rho_{local}=0.3$, $\rho_{merge}=0.2$, and $k$ chosen by~\eqref{para:k}. Where applicable, the nominal significance level is $\alpha=0.05$. Let $(\Sigma_0)_{jk}=0.5^{|j-k|}$, $\theta_j=\mathbf{1}\{j\leq\lfloor d/5\rfloor\}$, and define
\[
\boldsymbol{\mu}=\frac{5\delta}{4\log d}\theta,
\qquad
\Sigma_1=\left(1+\frac{2\sigma}{\sqrt d}\right)\Sigma_0.
\]
In Model~5, the distribution alternates between $\mathcal N_d(\mathbf 0,\Sigma_0)$ and $\mathcal N_d(\boldsymbol\mu,\Sigma_0)$, representing a normal mean change. In Model~6, the alternative is $\mathcal N_d(\boldsymbol\mu,\Sigma_1)$, so both mean and scale change. In Model~7, the base distribution is $\mathrm{Cauchy}_d(\mathbf 0,I_d)$ and the alternative is $\mathcal N_d(\boldsymbol\mu,\Sigma_1)$, representing a change from a heavy-tailed to a Gaussian regime.

Each sequence has $m=5$ change-points, starts with the base distribution, and alternates between the base and alternative distributions. The configurations are
\[
\begin{array}{lll}
\text{Well-separated:} & \mathcal T_{well}=\{70,140,210,280,350\}, & T=420,\\
\text{Frequent:} & \mathcal T_{freq}=\{170,185,200,215,230\}, & T=400,\\
\text{Nested:} & \mathcal T_{nested}=\{160,185,200,215,240\}, & T=400.
\end{array}
\]
In this order, the $(\delta,\sigma)$ values are $(2.8,0)$, $(4.8,0)$, and $(3.8,0)$ for Model~5; $(0.5,2.0)$, $(0.5,2.75)$, and $(0.5,2.75)$ for Model~6; and $(9.75,2.5)$, $(10.5,3.0)$, and $(10.0,3.0)$ for Model~7. These values yield moderately difficult problems. Because signal strengths vary across spacing configurations, comparisons between configurations reflect both spacing and signal strength. Each of the nine settings is repeated over 100 Monte Carlo replications.

We compare gBottomup with gMulti(G.WBS) and gMulti(G.SBS) \citep{zhang2021multi}, E-Divisive \citep{matteson2014nonparametric}, and KCP \citep{arlot2019kernel}. The latter two methods use the \texttt{R} package \texttt{ecp} \citep{james2015ecp}. The smallest spacing across the designs is 15. We set $\mathrm{MinLen}=10$ for both gMulti methods and the minimum cluster size for E-Divisive to $\lfloor\min_{0\leq j\leq m}(\tau_{j+1}-\tau_j)/2\rfloor$, with $\tau_0=0$ and $\tau_{m+1}=T$. For KCP, the maximum number of change-points is $2m$, and the penalty is calibrated under the simulation null to target a type~I error rate of 0.05. All other procedure-specific settings retain their defaults.

\subsection{Detection and segmentation results}\label{sec:na_result}

Table~\ref{tab:simulation_summary} reports the empirical frequency of estimating the correct number of change-points and the adjusted Rand index (ARI), which measures agreement with the true segmentation. The full distributions of $\widehat m-m$ are retained in Supplementary Tables~\ref{supp-tab:results_model5}--\ref{supp-tab:results_model7} in Supplement E.

\begin{table}[h]
\centering
\small
\setlength{\tabcolsep}{3pt}
\renewcommand{\arraystretch}{1.12}
\caption{Simulation results based on 100 replications per setting. Each entry gives the percentage of replications with $\widehat m=m$ followed by ARI. G.WBS and G.SBS denote gMulti(G.WBS) and gMulti(G.SBS), respectively.}
\label{tab:simulation_summary}
\begin{tabular}{@{}clccccc@{}}
\toprule
Model & Spacing & gBottomup & G.WBS & G.SBS & E-Divisive & KCP \\
\midrule
5 & Well-separated & 94 / .941 & 92 / .966 & 96 / .983 & 95 / .967 & 97 / .941 \\
  & Frequent       & 72 / .956 & 7 / .738  & 5 / .735  & 94 / .978 & 7 / .957 \\
  & Nested         & 64 / .935 & 14 / .712 & 18 / .720 & 90 / .971 & 0 / .819 \\
\midrule
6 & Well-separated & 70 / .943 & 8 / .813  & 33 / .873 & 0 / .099  & 0 / .000 \\
  & Frequent       & 69 / .937 & 11 / .779 & 17 / .812 & 0 / .697  & 0 / .000 \\
  & Nested         & 67 / .946 & 19 / .793 & 31 / .853 & 0 / .569  & 0 / .000 \\
\midrule
7 & Well-separated & 77 / .983 & 42 / .945 & 69 / .976 & 87 / .926 & 14 / .474 \\
  & Frequent       & 70 / .947 & 54 / .904 & 65 / .930 & 0 / .652  & 0 / .587 \\
  & Nested         & 68 / .946 & 44 / .895 & 57 / .928 & 0 / .561  & 0 / .556 \\
\bottomrule
\end{tabular}
\end{table}

For Model~5, all methods perform well when changes are well separated. E-Divisive also performs particularly well in the frequent and nested configurations, achieving higher exact-count frequencies and ARI than gBottomup. In those configurations, the graph-based top-down methods substantially underestimate the number of change-points, whereas gBottomup maintains ARI above 0.93 and substantially higher exact-count frequencies. KCP also underestimates the number of changes, despite a high ARI in the frequent configuration, illustrating the importance of assessing both count recovery and partition agreement.

For Model~6, gBottomup attains the highest ARI in all three configurations and estimates the correct number of change-points in 67--70\% of replications. The graph-based top-down methods frequently overestimate the number of changes in the well-separated configuration: the overestimation frequencies are 90\% for G.WBS and 63\% for G.SBS. Underestimation becomes more common in the frequent and nested configurations. E-Divisive underestimates across all three configurations, and KCP yields low segmentation accuracy under the settings used here.

For Model~7, gBottomup again attains the highest ARI across all configurations. The graph-based top-down methods remain competitive in ARI, but their count errors vary with the configuration, with overestimation particularly evident in the nested setting. E-Divisive performs well when changes are well separated but substantially underestimates their number in the frequent and nested configurations; KCP remains unstable. Taken together, the results show that gBottomup offers reliable performance across the three distributional scenarios, with advantages over the graph-based top-down alternatives for closely spaced changes and broader performance than competitors that excel primarily under pure mean changes.

%%%%%%%%%%%%%%%%%%%%%%%%%%%%%%%%%%%%%%%%%%%%%%%%%%%%%%%%%%%%%%%%%%%%%%%%%%%%%%

\section{Hydraulic system condition monitoring} \label{sec:real_data}

We evaluate gBottomup using condition monitoring data from an experimental hydraulic system \citep{helwig2015condition}. The system operates in repeated 60-second load cycles, with the conditions of four components varied across several severity levels: cooler efficiency, valve switching behavior, internal pump leakage, and accumulator gas pressure. The dataset contains measurements from multiple sensors over 2,205 cycles, including pressure, volume flow, motor power, vibration, and temperature. Our objective is to identify transitions between component conditions from the resulting sequence of high-dimensional sensor observations.

We restrict the analysis to cycles flagged as stable, retaining $T=1449$ cycles in their original order. Following \cite{mause2020detecting}, we focus on the six pressure sensors, each sampled at 100 Hz. Each 60-second cycle provides 6,000 measurements per sensor. The pressure measurements are standardized separately for each sensor. For each retained cycle, we concatenate the six standardized pressure profiles to form an observation $X_t\in\mathbb R^{36000}$. Thus, the analysis uses the full pressure profiles without additional feature extraction.

The recorded component conditions provide a reference segmentation: a reference change-point occurs between consecutive retained cycles when the condition of at least one component changes. This definition yields $m=143$ reference change-points. The component conditions follow a nested experimental ordering, with the cooler at the outermost level, followed by the accumulator, pump, and valve. We classify each reference change-point according to the outermost component whose condition changes. This gives 2 cooler-level, 9 accumulator-level, 24 pump-level, and 108 valve-level change-points. The classification allows us to assess how well each method recovers transitions at different levels of the experimental hierarchy.

For gBottomup, we use $L=5$ and $\alpha=0.05$, together with the tuning parameters selected in Section \ref{sec:choice_of_param}: $c=2$, $\rho_{local}=0.3$, and $\rho_{merge}=0.2$. The graph parameter $k$ is chosen according to (\ref{para:k}). We compare gBottomup with gMulti(G.WBS), gMulti(G.SBS), and E-Divisive. KCP is omitted because the null-model calibration used for its penalty in Section \ref{sec:analysis} cannot be transferred directly to this dataset. For E-Divisive, we set the minimum cluster size to $h=5$, half the minimum spacing of 10 cycles between reference change-points, following the rule used in Section \ref{sec:analysis}.

We assess performance using the ARI, the estimated number of change-points $\hat{m}$, and the number of recovered reference change-points, both overall and by component type. A reference change-point is counted as recovered if at least one estimated change-point lies within one cycle of it. All locations and matching tolerances are defined on the retained-cycle index.

\begin{table}[h]
\centering
\caption{Results for the hydraulic condition monitoring data ($T=1449$, $m=143$). The last five columns report the numbers of reference change-points recovered within a tolerance of one retained cycle, by component type and in total. Numbers in parentheses indicate the corresponding numbers of reference change-points.}
\label{tab:zema}
\begin{tabular}{lccccccc}
\hline
 & & & \multicolumn{5}{c}{Recovered reference change-points} \\
\cline{4-8}
Method & $\hat m$ & ARI & Valve (108) & Pump (24) & Acc.\ (9) & Cooler (2) & Total (143) \\
\hline
gBottomup     & 85 & \textbf{0.683} & \textbf{49} & \textbf{24} & \textbf{9} & \textbf{2} & \textbf{84} \\
gMulti(G.WBS) & 63 & 0.472 & 34 & 21 & 7 & 1 & 63 \\
gMulti(G.SBS) & 64 & 0.475 & 34 & 22 & 6 & 1 & 63 \\
E-Divisive    & 12 & 0.076 &  1 &  3 & 4 & 2 & 10 \\
\hline
\end{tabular}
\end{table}

Table \ref{tab:zema} summarizes the results. gBottomup achieves the highest ARI, 0.683, compared with 0.472 for gMulti(G.WBS), 0.475 for gMulti(G.SBS), and 0.076 for E-Divisive. It estimates 85 change-points and recovers 84 of the 143 reference boundaries. The two graph-based top-down methods each recover 63 reference boundaries, with 63 and 64 estimates, respectively. E-Divisive produces a much coarser segmentation, returning 12 estimates and recovering 10 reference boundaries. Thus, gBottomup achieves both greater boundary recovery and closer agreement with the reference partition.

\begin{figure}[h!]
	\centering
	\includegraphics[width=\columnwidth]{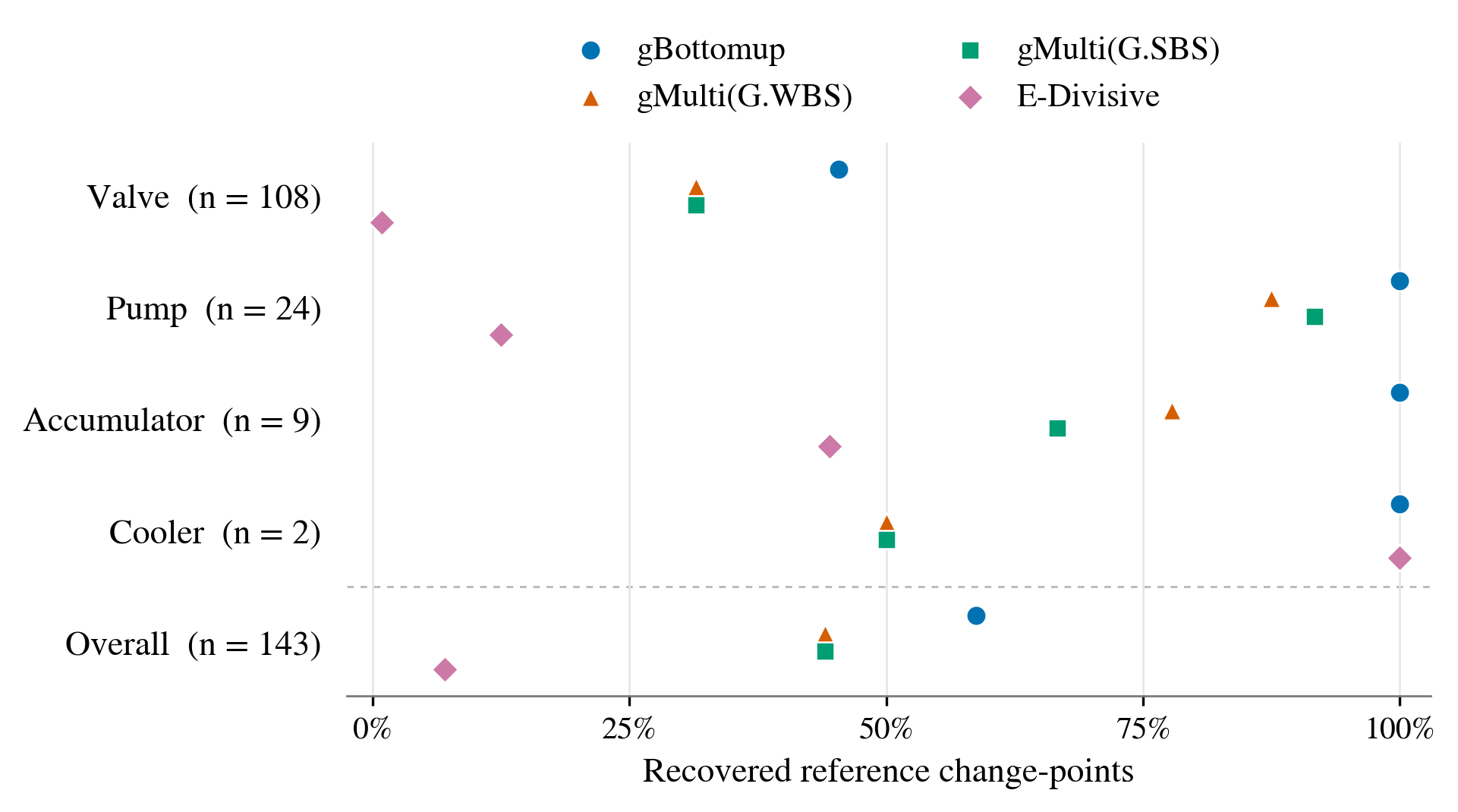}
	\caption{Proportion of reference change-points recovered in the hydraulic condition monitoring data, by component type and overall. Percentages are obtained from Table~\ref{tab:zema} using the reference counts shown in parentheses as denominators. A reference change-point is counted as recovered if an estimated change-point lies within one retained cycle of it. Symbols are offset vertically for clarity.}
	\label{fig:hydraulic_recovery}
\end{figure}

Figure \ref{fig:hydraulic_recovery} displays the component-specific recovery rates. gBottomup recovers all 35 pump-, accumulator-, and cooler-level change-points, whereas each graph-based top-down method recovers 29. For valve-level changes, the recovery rate is 45.4\% (49/108) for gBottomup, compared with 31.5\% (34/108) for each top-down method and 0.9\% (1/108) for E-Divisive. Of the ten reference boundaries recovered by E-Divisive, six correspond to accumulator- or cooler-level transitions.

Despite these improvements, even gBottomup recovers fewer than half of the valve-level boundaries. These transitions involve changes in the valve condition alone and occur at intervals of 10 cycles, leaving relatively few observations within each regime. The results support the usefulness of bottom-up scanning for recovering frequent changes in high-dimensional sensor data, while identifying the recovery of short valve-level regimes as a remaining challenge.

%%%%%%%%%%%%%%%%%%%%%%%%%%%%%%%%%%%%%%%%%%%%%%%%%%%%%%%%%%%%%%%%%%%%%%%%%%%%%%

\section{Conclusion}\label{sec-conc}

We proposed gBottomup for detecting multiple distributional changes in high-dimensional sequences. Local and hierarchical merging generate candidate boundaries, which are pruned using early-stopped ep-mBIC selection. The simulations demonstrate advantages over graph-based top-down alternatives for closely spaced changes and performance extending beyond pure mean changes. The hydraulic-system application further illustrates its usefulness for segmenting high-dimensional sensor records.

Although our implementation uses the generalized edge-count statistic \citep{chu2019asymptotic}, the bottom-up framework itself does not depend on this particular choice. In principle, any boundary score that is comparable across candidate windows and supports both candidate filtering and post-selection can be used in its place. This generality suggests natural extensions to other graph-based or nonparametric statistics, which we leave for future work.

Several directions remain for future work. On the theoretical side, the results in Section \ref{sec:theory} bound the Phase 2 evaluation budget and characterize the conservative pruning of ES-BE relative to selection over the same complete backward path. Statistical guarantees for candidate screening and final model selection, together with data-adaptive tuning, remain topics for future work. Further directions include data-adaptive tuning and extensions to alternative graph-based or nonparametric boundary scores, provided that they admit comparable calibration across windows and suitable candidate-filtering and model-selection criteria.

%%%%%%%%%%%%%%%%%%%%%%%%%%%%%%%%%%%%%%%%%%%%%%%%%%%%%%%%%%%%%%%%%%%%%%%%%%%%%%

\bibliographystyle{apalike}
\bibliography{bibliography.bib}

\end{document}